\documentclass[aps,prl,twocolumn,showkeys]{revtex4}
\usepackage{graphicx}
\usepackage{epsfig}
\usepackage{amssymb}
\usepackage{amsmath}
\usepackage{color}
\begin{document}

\title{Physical aging in article page views}
\author{Yeseul \surname{Kim}}
\thanks{These authors contributed equally to this article.}
\affiliation{Soft Matter Physics Laboratory, School of Advanced Materials Science and Engineering, SKKU Advanced Institute of Nanotechnology (SAINT), Sungkyunkwan University, Suwon 440-746, Korea}
\author{Kun \surname{Cho}}
\thanks{These authors contributed equally to this article.}
\affiliation{Soft Matter Physics Laboratory, School of Advanced Materials Science and Engineering, SKKU Advanced Institute of Nanotechnology (SAINT), Sungkyunkwan University, Suwon 440-746, Korea}
\author{Byung Mook \surname{Weon}}
\email{B.M.W. (bmweon@skku.edu)}
\affiliation{Soft Matter Physics Laboratory, School of Advanced Materials Science and Engineering, SKKU Advanced Institute of Nanotechnology (SAINT), Sungkyunkwan University, Suwon 440-746, Korea}

\date{\today}

\begin{abstract}
Statistics of article page views is useful for measuring the impact of individual articles. Analyzing the temporal evolution of article page views, we find that article page views usually decay over time after reaching a peak, especially exhibiting relaxation with nonexponentiality. This finding suggests that relaxation in article page views resembles physical aging as frequently found in complex systems.

\end{abstract}

\keywords{Statistics, Article page views, Physical aging, Relaxation}

\maketitle

\section{Introduction}

Nowadays, many scientific journals promote online publication, which is efficient in distributing research impact through web-based communities. Once a research article is published online on a journal website, the public can access the article immediately \cite{bib1}. Counting article page views is an alternative method to measure the impact of individual articles, instead of the journal impact factor \cite{bib2,bib3,bib4,bib5}. Empirically, when an article receives attention, its daily page views will reach a peak and then should decay with time, which is similar to a physical relaxation process. A recent study raised the possibility of a universal decay pattern in article page views, based on observations from six different PLoS journals \cite{bib1}. However, it remains unclear what mathematical model is appropriate for statistics in article page views.

Physical aging is the spontaneous temporal evolution of out-of-equilibrium systems \cite{bib6}. Glasses, for instance, usually show relaxation toward equilibrium, which is commonly referred to as physical aging \cite{bib7}. Nonexponential relaxation is ubiquitous in complex systems. Nonexponentiality in relaxation is frequently described by the stretched exponential \cite{bib8,bib9,bib10} or the Kohlrausch-Williams-Watts (KWW) decay function \cite{bib11,bib12} (also known as the Weibull function \cite{bib13}): $s = \exp[-(t/\alpha)^{\beta}]$, where the characteristic lifetime $\alpha$ corresponds to the specific lifetime for $s = \exp(-1)$ and the stretched exponent $\beta$ reflects the nonexponentiality. The simple exponential decay corresponds to $\beta = 1$ and the classical stretched exponential decay to $0 < \beta < 1$ (classically $\beta$ is invariant). Particularly, the stretched exponent is associated with a cascade mechanism of relaxation \cite{bib7}.

In this study, we utilize reliable statistical data of article page views that are available in Scientific Reports for articles published after 1 January 2012. Data for daily page views including HTML views and PDF downloads are offered 48 hours after online publication and updated daily. Here, we suggest a useful methodology for evaluating the nonexponentiality in article page views by adopting a modified stretched exponential function \cite{bib14,bib15,bib16,bib17,bib18}. This study shows the possibility that relaxation in article page views resembles physical aging as frequently found in complex systems.

\begin{figure}
\includegraphics[width=9cm]{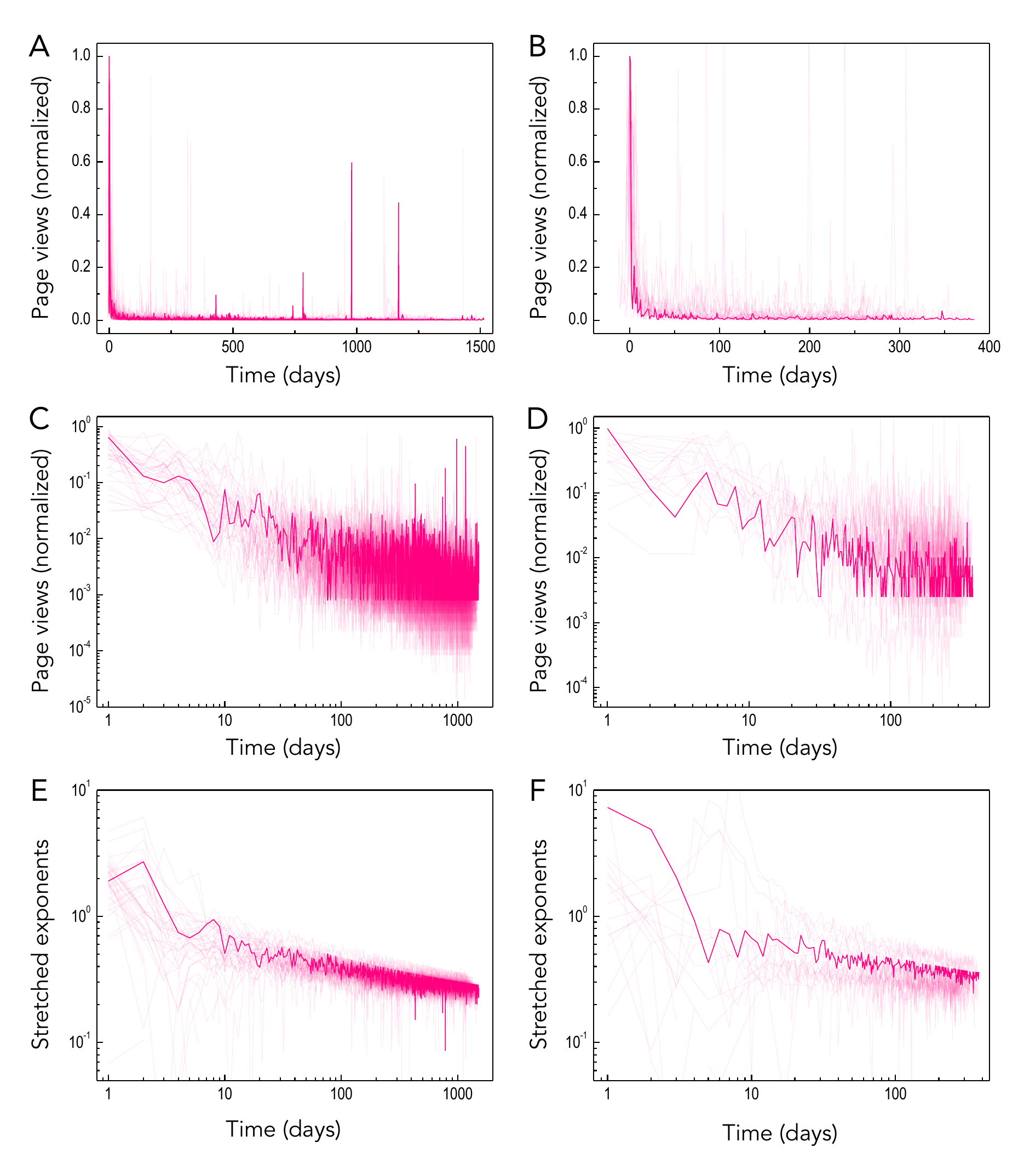}
\caption{{\bf Statistics of article page views after online publication.}
The first dataset (A,C,E) contains 42 articles selected from Scientific Reports published in 2012 and the second dataset (B,D,F) contains 19 articles published between January and May in 2015, showing more than 1,000 views at the early stage of online publication. The normalized page views over time are described by normal scales (A,B) and logarithmic scales (C,D). The temporal evolution of stretched exponents is obtained from the normalized page views, showing nonexponentiality in relaxation for both cases. As a guide, srep00223 (A,C,D) and srep07971 (B,E,F) are marked with bold solid lines.}
\label{fig1}
\end{figure}

\section{Results}

We selected two datasets: the first comprised 42 articles from Scientific Reports published in 2012 and the second comprised 19 articles published between January and May in 2015, showing more than 1,000 daily page views in a few days after online publication.

First, we counted the number of page views ($p_{t}$) at time $t$ (days) normalized by the number of initially maximized page views ($p_{0}$). The attention rate can be defined as $s = p_{t}/p_{0}$ and usually decreases with time from the initial peak. Empirically, there would be three types of public attention: initial attention at time $t_{0}$, irregular attention at time $t$, and no attention (for insignificant $p_{0}$). For demonstration, the normalized page views for the two datasets are illustrated by the normal scales in Fig~\ref{fig1}A-B. As illustrated, the maximum peak can exist at the origin of time, indicating the initial attention, and a few peaks can exist in the middle stage, indicating irregular attention.
To examine whether the normalized page views have power-law time dependencies, we depicted the log-log plots in Fig~\ref{fig1}C-D by simply rescaling Fig~\ref{fig1}A-B. Here the straightness in the log-log plots is not manifested, which means that the power-law scaling of $s$ vs $t$ is invalid. Such a temporal behavior in $s$ vs $t$ is repeatable for both datasets.

Next, we tested the nonexponentiality in the normalized page views. A convenient methodology is the adoption of a modified stretched exponential function. This function is described as $s = \exp[-(t/\alpha)^{\beta}]$ with the characteristic lifetime $\alpha$ and the stretched exponent $\beta$, especially where $\beta$ varies with time \cite{bib14,bib15}. This function was originally adopted to describe complicated biological survival curves \cite{bib16,bib17,bib18}. The stretched exponent is calculated as $\beta = \ln[-\ln(s)]/\ln(t/\alpha)$ from the $\alpha$ values. The characteristic lifetime $\alpha$ can be measured by detecting the interception point between $s(t)$ and $s(\alpha) = \exp(-1)$. The rough $\alpha$ estimate can be obtained from a linear regression from two data points that exist just above (+1p) and just below ($-$1p) the $\exp(-1)$ point for each $s$ vs $t$ curve \cite{bib18}. Here the stretched exponent is a good measure for testing the nonexponentiality of a decay curve and is relevant to an asymmetrical broadening of the relaxation time distribution \cite{bib9}.

As illustrated in Fig~\ref{fig1}E-F, we are able to see the nonexponentiality of the normalized page views for the two different datasets published in Scientific Reports. Interestingly, the stretched exponents behave irregularly over a short period (shorter than 10 days approximately) and then gradually decrease with time over a long period (longer than 10 days).

\section{Discussion}

Our finding from the page view statistics suggests that public attention in research is similar to physical relaxation processes in complex systems. Social media are indeed complex systems that actively connect people every day. Interestingly, the temporal evolution of the stretched exponents in the normalized page views is identical to that of luminescence decays: the stretched exponents become smaller than unity and gradually decrease with time (see refs. \cite{bib14,bib15}). Public attention and physical perturbation would be alike in terms of the time evolution of impact and information propagation. A stochastic model for information propagation would be relevant to the nonexponentiality of page view statistics \cite{bib1}. Further studies would be necessary on statistical analyses for a universal decay pattern in page views with wide datasets.

In summary, we present a novel statistical approach for article page views for online published articles. A significant finding is obtained: article page views usually decay over time after reaching a peak, especially exhibiting nonexponentiality. A feasible methodology is suggested for evaluation of the nonexponentiality. Our study shows that article page views follow nonexponential decay as public attention propagates through web-based communities, as frequently found in complex systems.

{\bf Acknowledgments}

This work was supported by Sungkyun Research Fund, Sungkyunkwan University, 2014.

\end{document}